\documentclass[aps,prb,superscriptaddress,showpacs,floatfix]{revtex4-1}
\usepackage{graphicx,amsmath,amssymb,xspace,epsfig,float,multirow,subfigure,tabularx}

\pdfoutput=1

\begin{document}

\title{Concurrent intersection point of magnetization and
magneto-conductivity curves in strongly fluctuating superconductors}
\author{Cheng Chi}
\affiliation{School of Physics, Peking University, Beijing 100871, \textit{China}}
\affiliation{Collaborative Innovation Center of Quantum Matter, Beijing, China}
\author{Xujiang Jiang}
\affiliation{School of Physics, Peking University, Beijing 100871, \textit{China}}
\affiliation{Collaborative Innovation Center of Quantum Matter, Beijing, China}
\author{Jiangfan Wang}
\affiliation{School of Physics, Peking University, Beijing 100871, \textit{China}}
\affiliation{Collaborative Innovation Center of Quantum Matter, Beijing, China}
\author{Dingping Li}
\email{lidp@pku.edu.cn}
\affiliation{School of Physics, Peking University, Beijing 100871, \textit{China}}
\affiliation{Collaborative Innovation Center of Quantum Matter, Beijing, China}
\author{Baruch Rosenstein}
\email{baruchro@hotmail.com}
\affiliation{Electrophysics Department, National Chiao Tung University, Hsinchu 30050,
\textit{Taiwan, R. O. C}}
\date{\today }

\begin{abstract}
The thermal fluctuations contribution to magnetization and
magneto-conductivity of type II layered superconductor is calculated in the
framework of Lawrence-Doniach model. For numerous high temperature cuprate
superconductors, it was discovered that the magnetization dependence on
temperature in wide range of fields exhibits an intersection point at a
temperature slightly below $T_{c}$. We notice a similar intersection point
of the magneto-conductivity curves at the approximate same temperature. The
phenomenon is explained by strong (non-gaussian) thermal fluctuations with
interactions treated using a self-consistent theory. All higher Landau
levels should be included. Dimensionality of the fluctuations is defined and
the 2D-3D dimensional crossover is the key for the existence of intersection
points.
\end{abstract}

\pacs{74.20.De, 74.25.Bt, 74.25.Ha, 74.40.-n}
\maketitle

\section{Introduction}

Discovery of high temperature superconductors (HTSC) attracted attention to
effects of thermal fluctuations on the thermodynamic, magnetic and transport
of type-II superconductors. In these materials, even at zero magnetic field,
the role of thermal fluctuations is enhanced by several factors including
high critical temperature, short coherence length and large anisotropy. Very
large second critical magnetic field makes the highest available fields
accessible for experiments in the superconducting states. Strong magnetic
field greatly enhances the effect of the fluctuations making the magnetic
phase diagram very complicated by creating the vortex liquid state over
large range of temperatures below $T_{c}$ (clearly seen in magnetization\cite%
{Zeldov1995} and specific heat\cite{Schilling1997,Junod1999}), and
broadening the resistance drop in magneto-resistence\ upon transition.

More recently the fluctuations effects well above $T_{c}$ have been studied
in detail for magnetization\cite{LuLi2010}, electric conductivity\cite%
{Rullier2011}, and Nernst effect\cite{Xu2000,Wang2006}. The magnitude of
thermal fluctuation is quantified by the Ginzburg number $Gi$,\ which can
reach $10^{-2}$-$10^{-1}$ in high-$T_{c}$ cuprate superconductors in
contrast with $10^{-9}$-$10^{-6}$ for conventional low-$T_{c}$
superconductors. In other unconventional superconductors like pnictides the
fluctuations are still very significant. Though the multi-band structure of
the iron-based superconductors is different from cuprate superconductors,
they show many similarities like the two dimensional layered-structure.

If a superconductor is strongly fluctuating, ``virtual" or ``preformed"
Cooper pairs exist above $T_{c}$, but the order parameter $\Psi $ is not
phase coherent. Average of its amplitude (related to the superfluid density-
$\left \langle \left \vert \Psi \right \vert ^{2}\right \rangle $)\ might be
sufficiently large to dominate electromagnetic properties like magnetization
and magneto-transport over the typically small normal background.

When magnetization was measured, it was found surprisingly that, when the
magnetization as a function of temperature, $M\left( T\right) $, plotted at
different magnetic fields $H$, the curves intersect at the same temperature $%
T^{\ast }$. The first clear demonstration of this observation in $%
YBa_{2}Cu_{3}O_{7-\delta }$ ($YBCO$) \cite{Welp1991} and $%
Bi_{2}Sr_{2}CaCu_{2}O_{8}$ ($BSCCO$) \cite{Kes1991} has been made in early
nineties, and later extended to $Tl_{2}Ba_{2}CaCu_{2}O_{8}$ ($TBCCO$) \cite%
{Wahl1997}, $HgBa_{2}Ca_{2}Cu_{3}O_{8+\delta }$ ($HgBCCO$) \cite%
{Naughton2000} and $La_{2-x}Sr_{x}CuO_{4+\delta }$ ($LSCO$) \cite%
{Finnemore2002}. The physics in the relevant part of the magnetic phase
diagram is that of the vortex liquid. The basic vortex liquid theory was
developed in eighties\cite{Thouless1975,Thouless1976} based on the idea of a
homogeneous state with finite correlation determined by the energy gap $%
\varepsilon $ characterizing short range order. Typically the energy gap is
estimated theoretically in a self - consistent manner.

Over the years there have been several attempts to explain the appearance of
the intersection point of the magnetization curves. First the theory was
restricted (due to complexity of a nonperturbative problem) to the lowest
Landau level (LLL). The restriction allowed to obtain nonperturbative
expressions even in the strongly fluctuating cases both in 2D and 3D, see%
\onlinecite{Tesanovic1994} and references therein. Bulaevskii et al\cite%
{Bulaevskii1992} attempted to use the 2D version of the\ theory to explain
the intersection point in $YBCO$, however it was subsequently realized that
the exact intersection is inconsistent with the strict LLL scaling\cite%
{Rosenstein2005}.

When the restriction on the first Landau level was lifted\cite{Jiang2014}
(while still retaining the self-consistent fluctuation theory of the vortex
liquid), the Ginzburg Landau (GL) theory extended to layered materials
became capable to describe magnetization in $LSCO$, $BSCCO$ and $YBCO$ in
wide range of fields and temperatures (even above $T_{c}$) with small number
of parameters. Both experimentally and theoretically it became apparently
that the intersection point is never exact. It depends slightly on magnetic
field in surprisingly wide range of fields, but beyond this range the
phenomenon quickly disappears.

An interesting question arises whether the similar \textquotedblleft
intersection point" also appears in other fluctuation phenomena like
fluctuation transport, for example, the magneto-resistance. Ullah and Dorsey%
\cite{Dorsey1991} obtained expressions for the scaling behavior as a
function of magnetic field and temperature of various thermodynamic and
transport quantities within the self-consistent Hartree approximation and
under the LLL restriction. They demonstrated that the product of the
superconducting part of conductivity $\sigma _{s}$ (due to the order
parameter) and magnetic field, \thinspace $K\equiv \sigma _{s}H$, scales
exactly as the magnetization $M$ in the LLL restriction. The 3D LLL scaling
in part of magnetic phase diagrams was later confirmed experimentally\cite%
{Welp1991} in optimally doped $YBCO$ and later in other cuprates\cite%
{Lan1993,Kim1992} and pnictides\cite{Pallecchi2009,Liu2010}. As a
consequence, one would expect intersection points for $K$. We therefore have
replotted $K$ as function of temperature for $YBCO$\cite{Palstra1989}, see
Fig.1, and one iron pnictide superconductor, $LaFeAsO_{0.9}F_{0.1-\delta }$%
\cite{Chen2008}discussed below. The curves intersect approximately at the
same temperature $T^{\ast }$ for different magnetic fields $H$.

As the vortex pinning is the  mechanism for the existing of superconducting
states in type II superconductors and therefore it might can not be ignored.
However the intersection point region is close to the critical temperature,
and in highly fluctuating superconductors (large Ginzburg number), due to
strong thermal fluctuation the pinning effect for vortex liquid is very
small due to thermal de-pinning. Therefore the pinning and the pair breaking
scattering due to pinning can be ignored in this paper.

\begin{figure}[tbp]
\begin{center}
\includegraphics[width=8cm]{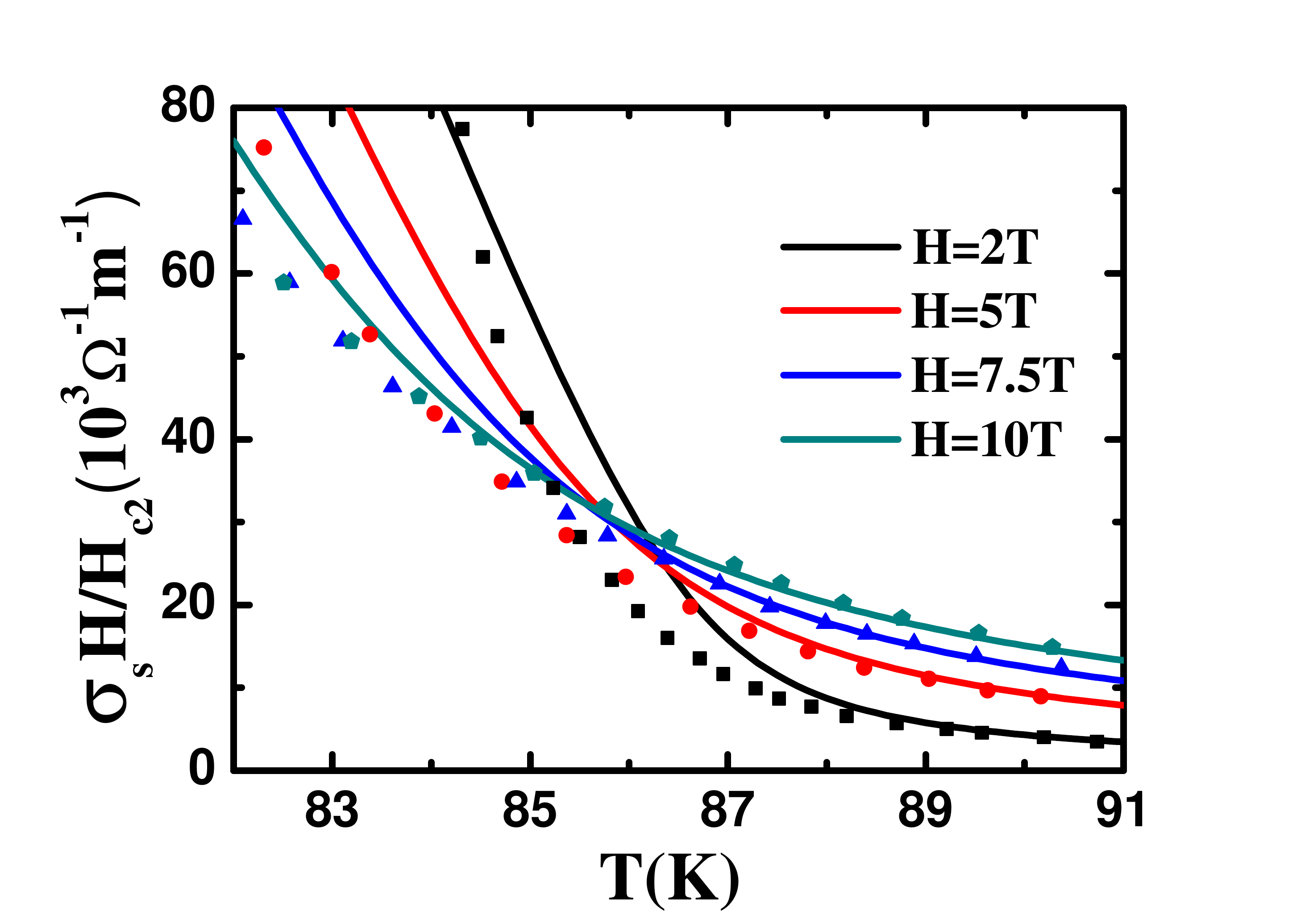}
\end{center}
\par
\vspace{-0.5cm}
\caption{The intersection points of conductivity times magnetic field vs
temperature $T$ of $YBCO$.}
\end{figure}

In this paper we present a quantitative theory of magnetization and magneto
- conductivity and explain the intersection points within the
phenomenological GL framework. It turns out that the layered structure,
determining the dimensionality of the thermal fluctuations, is crucial for
the appearance of the concurrent intersection points of various physical
quantities, so the Lawrence Doniach (LD) model should be used. The
calculation of conductivity requires the time dependent version of the
model. The same sufficiently precise self - consistent method in the vortex
liquid phase should be used to obtain simultaneously magnetization and
conductivity.

The formulas of the magnetization and conductivity are used to fit the
experimental data of various materials. It turns out that $T^{\ast }$ is
located in the 2D - 3D crossover temperature regime in which the coherence
length $\xi _{c}\left( T\right) $ in the direction perpendicular to the
layers is roughly equal to the interlayer spacing.

The paper is organized as follows. The theoretical calculation of
fluctuation magnetization and conductivity based on the LD model and time
dependent Ginzburg Landau (TDGL) theory are introduced in Sec. II. The
intersection points of fluctuation magnetization is calculated in Sec. III.
The intersection points of conductivity curves of cuprate superconductors
and pnictide superconductors are discussed in Sec. IV. We conclude in Sec.
V. that the intersection points of magnetization and conductivity are due to
the 2D-3D dimensional crossover.

\section{Fluctuation magnetization and conductivity in the Lawrence -
Doniach model.}

\subsection{Model}

The Lawrence-Doniach model are used to study the vortex matter in layered
superconductors in the Ginzburg - Landau phenomenological approach. The
Ginzburg - Landau free energy is expressed in terms of complex order
parameter $\Psi _{n}\left( \mathbf{r}\right) $ in the $n$-th superconducting
layer:%
\begin{eqnarray}
F_{GL} &=&d^{\prime }\sum_{n}\int d\mathbf{r}\left[ \frac{\hbar ^{2}}{%
2m^{\ast }}\left \vert \mathbf{D}\Psi _{n}\right \vert ^{2}+\frac{\hbar ^{2}%
}{2m_{c}d^{\prime 2}}\left \vert \Psi _{n}-\Psi _{n+1}\right \vert
^{2}\right.  \notag \\
&&\left. +\alpha \left( T-T_{\Lambda }\right) \left \vert \Psi _{n}\right
\vert ^{2}+\frac{\beta }{2}\left \vert \Psi _{n}\right \vert ^{4}\right]
\text{.}  \label{2-1}
\end{eqnarray}%
Here $\mathbf{r}$ is position in the superconducting plane. The second term
describes the Josephson coupling between the planes separated by the
inter-layer distance $d^{\prime }$. Applied field is assumed to be
perpendicular to the planes and much larger that the lower critical field $%
H_{c1}$ so that in the vortex liquid phase the magnetic induction $B$ is
homogeneous and magnetization is small compared to it. Therefore the vector
potential in Landau gauge in the covariant derivative, $\mathbf{D=\nabla +}%
i\left( 2e/\hbar c\right) \mathbf{A}$\textbf{,} can be approximated by $%
\mathbf{A=}\left( -By,0\right) $.

Parameters are as follows. The effective mass of a Cooper pair in the $a$-$b$
plane is $m^{\ast }$, while the one along the $c$ axis is $m_{c}$, so that
the anisotropy parameter is $\gamma =\sqrt{m_{c}/m^{\ast }}$. The mean-field
critical temperature in Eq.(\ref{2-1}) is denoted by $T_{\Lambda }$ to
stress its dependence on the ultraviolet cutoff $\Lambda $ (of the order of
inverse lattice spacing). Thermal fluctuations of the order parameter on the
mesoscopic scale, described by a Boltzmann sum, are generally characterized
by the dimensionless 3D Ginzburg number,
\begin{equation}
Gi=\frac{1}{2}\left( \frac{8e^{2}\kappa ^{2}\xi _{\Lambda }T_{c}\gamma }{%
\hbar ^{2}c^{2}}\right) ^{2}\text{.}  \label{Gi}
\end{equation}%
The Ginzburg-Landau parameter $\kappa $ is the ratio of the penetration
depth, $\lambda _{\Lambda }=\frac{c}{2e^{\ast }}\sqrt{\frac{m^{\ast }\beta }{%
\pi \alpha T_{\Lambda }}}$, and the in - plane coherence length, $\xi
_{\Lambda }=\hbar /\sqrt{2m^{\ast }\alpha T_{\Lambda }}$, and $T_{c}$ is the
critical temperature.

To study the transport properties of a layered superconductor, the time
dependent Ginzburg Landau Lawrence-Doniach model is used in a magnetic field
near the mean-field transition temperature\cite{Hohenberg1977}:%
\begin{equation}
\frac{\hbar ^{2}\gamma _{D}}{2m}D_{\tau }\Psi _{n}=-\frac{1}{d^{\prime }}%
\frac{\delta F_{GL}}{\delta \Psi _{n}^{\ast }}+\zeta _{n}\text{.}
\label{TDGL}
\end{equation}%
Here $D_{\tau }\equiv \partial /\partial _{\tau }-i\left( e^{\ast }/\hbar
\right) \phi $ is the covariant time derivative with $\phi =-Ey$ being the
scalar potential describing the electric field $E$ applied along the $y$
direction. The thermal noise term $\zeta _{n}$ should satisfy the
fluctuation-dissipation theorem and ensure that the system relaxes to the
proper equilibrium distribution,%
\begin{equation}
\left \langle \zeta _{m}^{\ast }\left( \mathbf{r},\tau \right) \zeta
_{n}\left( \mathbf{r}^{\prime },\tau ^{\prime }\right) \right \rangle =\frac{%
\hbar ^{2}\gamma _{D}}{m^{\ast }d^{\prime }}T\delta _{m,n}\delta \left(
\mathbf{r-r}^{\prime }\right) \delta \left( \tau -\tau ^{\prime }\right)
\text{,}
\end{equation}%
where the $\left \langle \cdots \right \rangle $ denotes a thermal average.

\subsection{Static properties of the vortex liquid phase within the self -
consistent approximation.}

The self - consistent fluctuation approximation (SCFA) in statics has been
used to derive the magnetization in the vortex liquid phase in Ref. %
\onlinecite{Jiang2014} in which the full derivations of the magnetization
formula used below can be found. The basic characteristics of the vortex
liquid is the excitation energy gap, which will be denoted by $\varepsilon $
in physical energy unit $\frac{2\hbar eH_{c2}}{m^{\ast }c}$($H_{c2}=\hbar
c/2e\xi _{\Lambda }^{2}$ is the upper critical field). It is determined by
the gap equation

\begin{eqnarray}
\varepsilon &=&-a_{h}-\frac{\omega t}{\pi }\ln \left( 1+\Lambda d^{2}+\sqrt{%
2\Lambda d^{2}+\Lambda ^{2}d^{4}}\right)  \notag \\
&&+\frac{\omega td}{2\pi ^{2}}\int_{k_{z}=0}^{2\pi /d}\left[ \psi \left(
g\left( k\right) +\Lambda /b\right) -\psi \left( g\left( k\right) \right) %
\right] \text{,}  \label{epsilon}
\end{eqnarray}%
where $k$ is the wave vector along the magnetic field direction and $\Lambda
$ is the dimensionless cutoff energy in physical energy unit. The
dimensionless ``distance" from the $H_{c2}$ line is $a_{h}=\left(
1-t-b\right) /2$, where $t=T/T_{c}$, $b=B/H_{c2}$ . The parameter $\omega $
describes the thermal fluctuation strength of the layered superconductors
often expressed via the Ginzburg number, Eq.(\ref{Gi}),

\begin{equation}
\omega =\frac{\pi }{d}\sqrt{2Gi}\text{.}
\end{equation}%
where $\Gamma $ and $\psi $ are the gamma and the digamma functions
respectively and dimensionless layer distance is $d=d^{\prime }\gamma /\xi
_{\Lambda }$. It is convenient to introduce a function of the perpendicular
wave vector $k$ frequently used in the equations below,%
\begin{equation}
g\left( k\right) =\frac{1}{b}\left( \frac{1-\cos kd}{d^{2}}+\varepsilon
\right) \text{.}
\end{equation}

The critical temperature $T_{c}$ is often smaller than the meanfield
critical temperature $T_{\Lambda }$ due to strong thermal fluctuations on
the mesoscopic scale\cite{Jiang2014}. Within the gaussian approximation the
relation is given by
\begin{equation}
T_{c}=T_{\Lambda }\left[ 1-\frac{2\omega }{\pi }\ln \left( 1+\Lambda d^{2}+d%
\sqrt{2\Lambda +\Lambda ^{2}d^{2}}\right) \right] \text{.}
\end{equation}

Magnetization was calculated in the framework of SCFA including all Landau
levels\cite{Jiang2014}:%
\begin{eqnarray}
M &=&-\frac{2eT}{hcd^{\prime }}\left \{ \frac{\Lambda }{b}+\frac{d}{2\pi }%
\int_{k}\left[ \ln \Gamma \left( g\left( k\right) +\frac{\Lambda }{b}\right)
\right. \right.  \notag \\
&&-\left( g\left( k\right) +\frac{\Lambda }{b}-\frac{1}{2}\right) \psi
\left( g\left( k\right) +\frac{\Lambda }{b}\right)  \notag \\
&&\left. \left. +\left( g\left( k\right) -\frac{1}{2}\right) \psi \left(
g\left( k\right) \right) -\ln \Gamma \left( g\left( k\right) \right) \right]
\right \} \text{.}  \label{mag}
\end{eqnarray}%
Dynamical properties of the vortex liquid require the time dependent
equation Eq.(\ref{TDGL}) involving an extra parameter, the diffusion
constant $\gamma _{D}$.

\subsection{Conductivity within the self - consistent approximation.}

While for a BCS superconductor the diffusion constant $\gamma _{D}$ is
related to parameters in GL model by $\gamma _{BCS}=\frac{\pi \hbar }{%
8T_{\Lambda }\xi _{\Lambda }^{2}}$, for unconventional superconductors, this
relation can be modified, $\gamma _{D}=\eta \gamma _{BCS}$ ($\eta $ is a
fitting parameter of order $1$).

The magneto conductivity of the layered superconductor due to
superconducting fluctuation in the vortex liquid phase using SCFA was
studied in Refs. \onlinecite{Bui2010,Wang2016}, including high Landau
levels. The Cooper pair contribution to the conductivity in terms of the
excitation energy $\varepsilon $, determined by the gap equation Eq.(\ref%
{epsilon}), is
\begin{eqnarray}
\sigma _{s} &=&\frac{e^{2}t\eta \gamma }{8hb\xi _{\Lambda }}\int_{k}\left \{
\left( 2g\left( k\right) -1\right) \left[ \psi \left( g\left( k\right)
\right) +\psi \left( g\left( k\right) +\frac{\Lambda }{b}+\frac{1}{2}\right)
\right. \right.  \notag \\
&&\left. \left. -\psi \left( g\left( k\right) +\frac{1}{2}\right) -\psi
\left( g\left( k\right) +\frac{\Lambda }{b}\right) \right] +\frac{\Lambda }{%
\Lambda -b/2+g\left( k\right) b}\right \} \text{.}  \label{con}
\end{eqnarray}%
The detailed derivation of Eq.(\ref{con}) can be found in %
\onlinecite{Wang2016}. Having expressed both the static and the dynamical
physical quantities within the same approximation, we now can turn to the
main point of the present study: the intersection points at different
magnetic fields. Let us start with magnetization.

\section{Intersection points of magnetization curves}

As mentioned in Introduction, the intersection points for the magnetization,
defined as,%
\begin{equation}
\left. \frac{\partial M}{\partial H}\right \vert _{T=T^{\ast }}=0\text{,}
\label{3-1}
\end{equation}%
were measured in many high $T_{c}$ cuprates \cite%
{Salem09,Rosenstein2001,Naughton2000,Finnemore2002} and explained within the
``lowest Landau level" approximation \cite{Bulaevskii1992,Rosenstein2005}.
To calculate the intersection curve determined by Eq.(\ref{3-1}) from Eq.(%
\ref{mag}), the derivative of magnetization with respect to the magnetic
field is required,%
\begin{eqnarray}
\frac{\partial M}{\partial b} &=&-\frac{eT\gamma }{\pi hc\xi _{\Lambda }b}%
\int_{k}\left[ -\left( g\left( k\right) +\frac{\Lambda }{b}-\frac{1}{2}%
\right) \left( \varepsilon _{b}^{\prime }-g\left( k\right) -\frac{\Lambda }{b%
}\right) \psi ^{\prime }\left( g\left( k\right) +\frac{\Lambda }{b}\right)
\right.  \notag \\
&&\left. +\left( g\left( k\right) -\frac{1}{2}\right) \psi ^{\prime }\left(
g\left( k\right) \right) \left( \varepsilon _{b}^{\prime }-g\left( k\right)
\right) -\frac{\Lambda }{b}\right] \text{.}
\end{eqnarray}%
Here the derivative $\varepsilon _{b}^{\prime }=\partial \varepsilon
/\partial b$, that can be calculated using the gap equation, Eq.(\ref%
{epsilon}), and the intersection curve $T^{\ast }\left( H\right) $ is
obtained numerically.

\begin{figure}[tbp]
\begin{center}
\includegraphics[width=8cm]{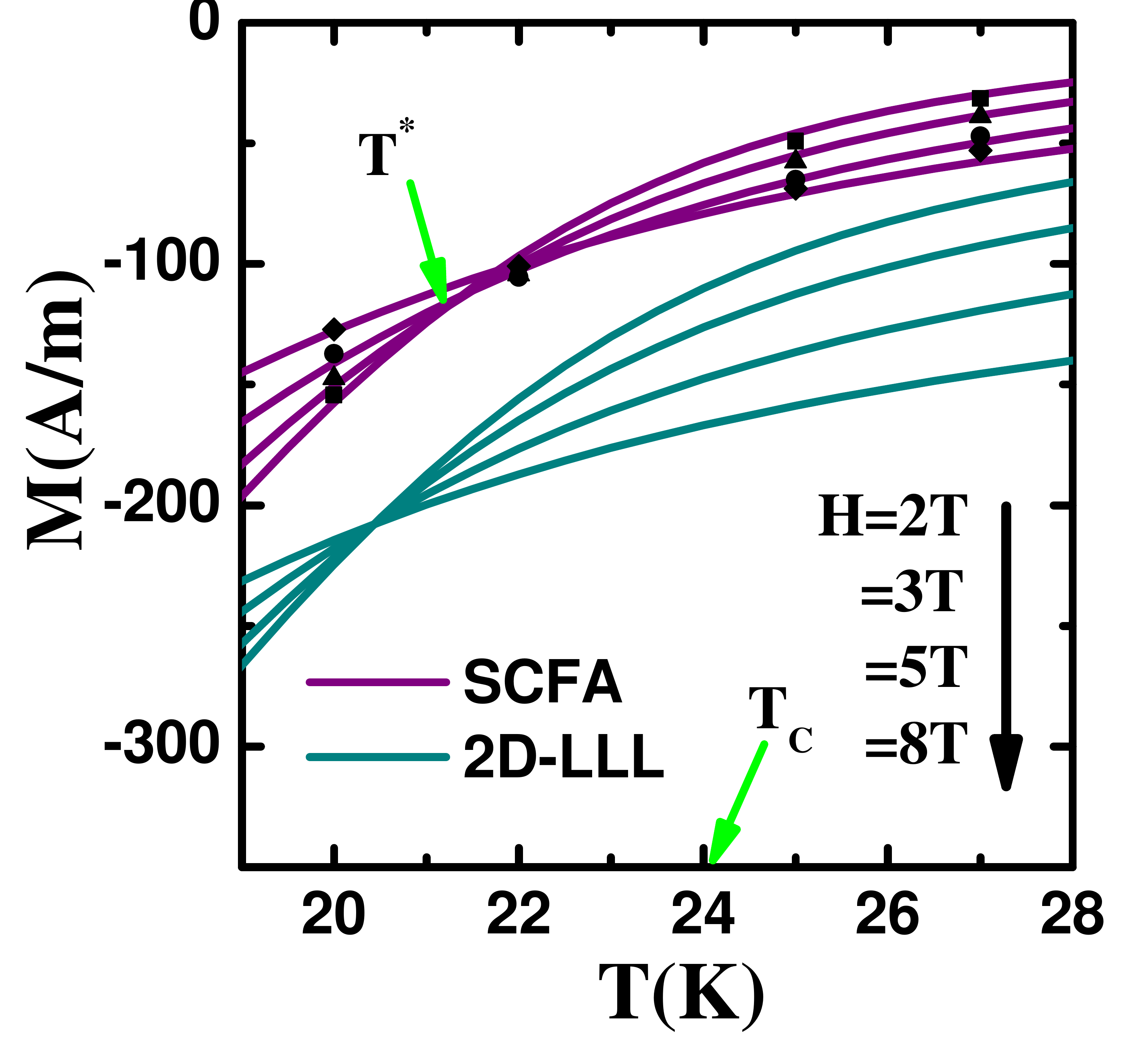}
\end{center}
\par
\vspace{-0.5cm}
\caption{Magnetization data of Ref. \onlinecite{LuLi2010} and theoretical
fits(purple lines) of $LSCO$ crystal for various values of $H$. The
comparison between the fluctuation magnetization calculated using the
self-consistent fluctuation approximation (SCFA) vs the 2D lowest Landau
level(2D-LLL) one.}
\end{figure}

\bigskip

In Fig. 2 magnetization curves of the under\ - doped$\
La_{1.91}Sr_{0.09}CuO_{4}$ ($T_{c}=24K$, $d^{\prime }=6.58\mathring{A}$) at $%
H=2T$, $3T$, $5T$, $8T$ fitted by Eq.(\ref{mag}) (purple lines) are shown.
The fitting parameters\cite{Jiang2014} are $H_{c2}=31T$, $\gamma =29$, $%
\Lambda =0.3$, $Gi=0.033$ and $\omega =0.138$. The result of SCFA
approximation with all Landau level approach fits well with experiments
(points). The magnetization in 2D-LLL approximation (detailed derivations
are presented in Appendix A) is also shown in Fig. 2 (green lines). The
2D-LLL approximation\ theory predicts a single intersection point in
contrary to the experimental data. Furthermore, it gave too big
diamagnetization (roughly twice) compared to the SCFA result.

\begin{figure}[tph]
\begin{center}
\includegraphics[width=8cm]{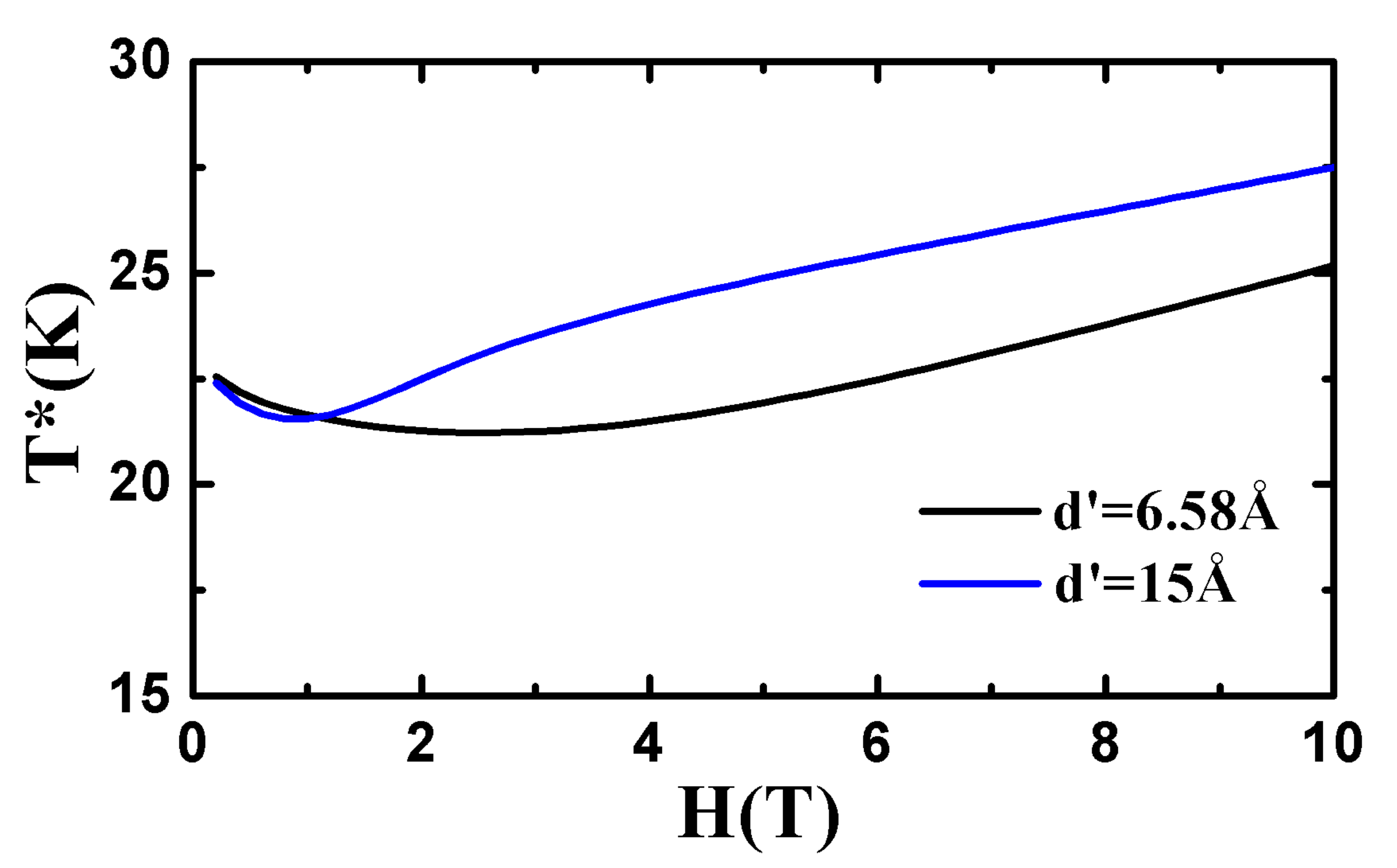}
\end{center}
\par
\vspace{-0.5cm}
\caption{The intersection point lines $T^{\ast }\left( H\right) $ of
magnetization vs magnetic fields $H$ for different interlayer spacing $%
d^{\prime }$ of $LSCO$ crystal.}
\end{figure}

In Fig.3, the intersection point lines $T^{\ast }\left( H\right) $ of two
superconductors with different interlayer spacings are given. The first is
underdoped $LSCO$ with $d^{\prime }=6.58\mathring{A}$ while the second
(hypothetical) has $d^{\prime }=15\mathring{A}$ and all the other
parameters, $T_{c}$, $H_{c2}$, $Gi$, $\gamma $, the same. For the first
material, $T^{\ast }$ is nearly independent of magnetic field in the range $%
1T$ to $5T$ ($T^{\ast }\left( 1T\right) =21.64K$, $T^{\ast }\left( 2T\right)
=21.27K$, $T^{\ast }\left( 3T\right) =21.25K$, $T^{\ast }\left( 4T\right)
=21.5\ K$, $T^{\ast }\left( 5T\right) =21.93K$). The best intersection point
is at $21.4K$. For the hypothetical more anisotropic material, $T^{\ast }$
exhibits stronger dependence on $H$. Therefore in the $2D$ limit, there is
no well defined intersection point. Since the earlier explanation of the
intersection point\cite{Bulaevskii1992,Rosenstein2005} made use of 2D limit
and LLL approximation, let us clarify why it is not likely.

\begin{figure}[tph]
\begin{center}
\includegraphics[width=8cm]{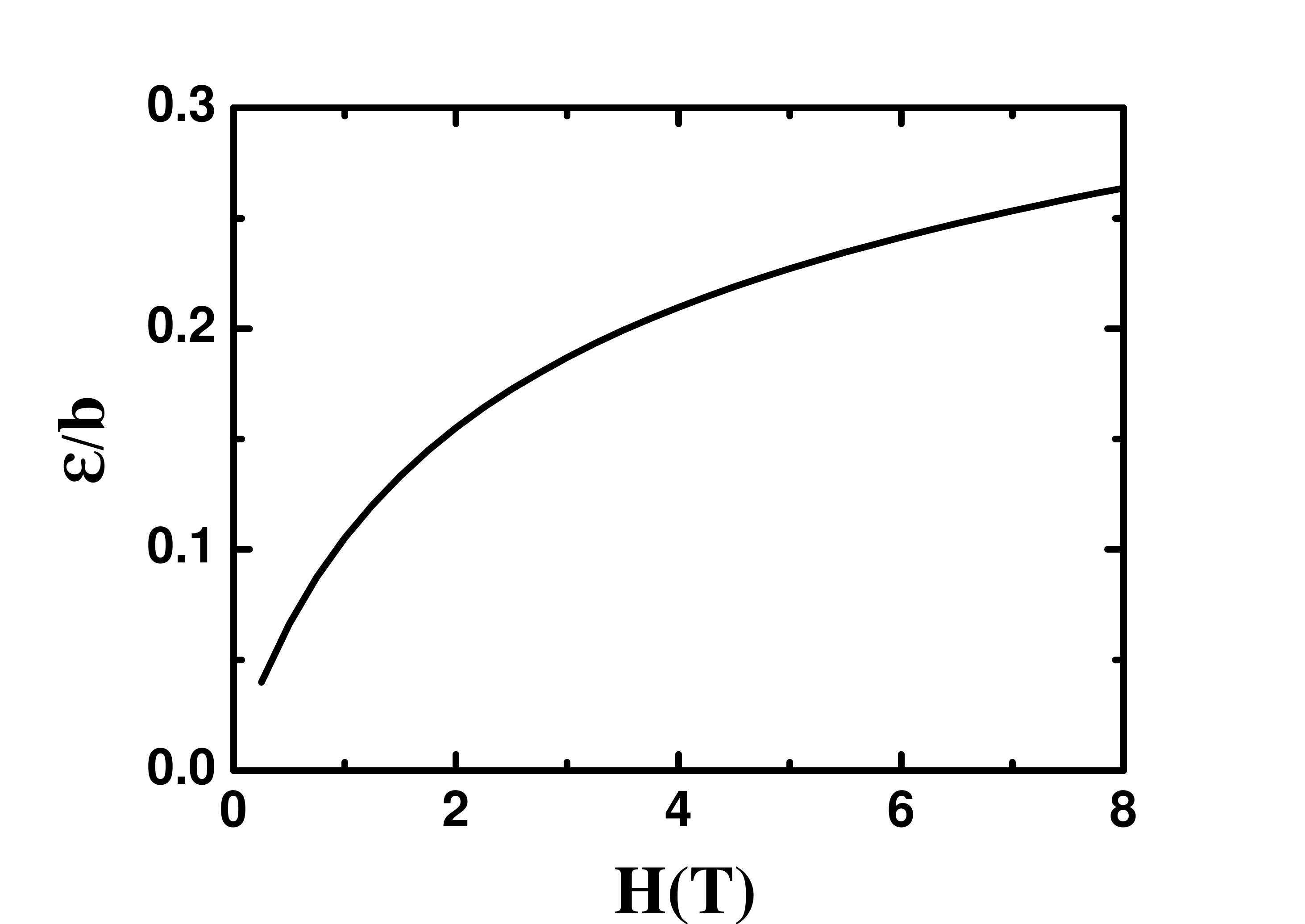}
\end{center}
\par
\vspace{-0.5cm}
\caption{ The $\protect \varepsilon \left( b\right) /b$\ as function of
magnetic field with fixed $d^{\prime }$=6.58\AA \ and $T^{\ast }=21.4$K of $%
LSCO$ crystal.}
\end{figure}

In Fig.4 the energy gap of the vortex liquid, $\varepsilon \left( b\right)
/b $,\ in underdoped $LSCO$ at $T^{\ast }=21.4K$ is given as a function of
magnetic field. The LLL approximation condition $\varepsilon \left( b\right)
<<b$ is questionable since $\varepsilon \left( b\right) /b$ exceeds $0.25$.
So one has to look for an explanation elsewhere. An alternative is the
dimensional 2D - 3D crossover taking place when coherence length in
direction perpendicular to layers becomes comparable to the interlayer
distance.

\begin{figure}[tph]
\begin{center}
\includegraphics[width=8cm]{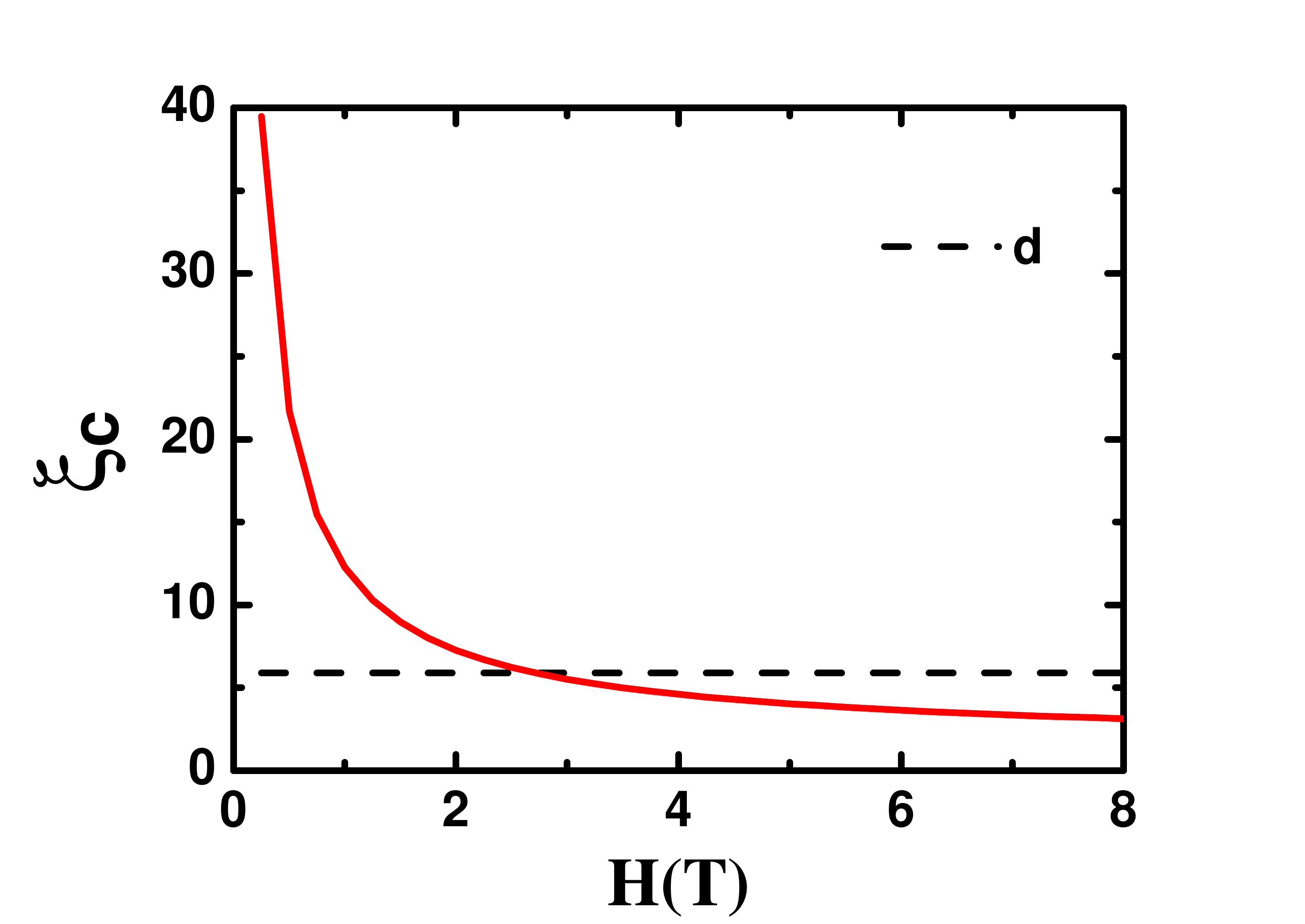}
\end{center}
\par
\vspace{-0.5cm}
\caption{$\protect \xi _{c}\left( H\right) $ as the function of magnetic
field with fixed $d^{\prime }=6.58$\AA \ ($d=5.87$) and $T^{\ast }=21.4$K of $%
LSCO$ crystal.}
\end{figure}

The correlation length $\xi _{c}$ is calculated in the Appendix. B. In
Fig.5, it is shown for underdoped $LSCO$ at $T^{\ast }$ as a function of
magnetic field. One notices that $\xi _{c}\left( H\right) =d^{\prime }$ at $%
H=2.7T$ just at the point in which the intersection point is defined the
best (see a minimum in Fig.3). To conclude, the intersection points lies
near the 2D - 3D crossover. This is one of the main results of the present
paper.

The quantity $K$ shows also intersection point as it will be shown below.
The definition of the intersection point for this quantity is%
\begin{equation}
\left. \frac{\partial }{\partial H}K\right \vert _{T=T^{\ast }}=0\text{.}
\label{4-1}
\end{equation}%
The derivative of $K$ with respect to $b$ simplifies to%
\begin{eqnarray}
\frac{\partial K}{\partial b} &=&\frac{et\eta \gamma c}{32\pi b\xi _{\Lambda
}^{3}}\int_{k}\left \{ 2\left( \varepsilon _{b}-g\left( k\right) \right) %
\left[ \psi \left( g\left( k\right) +\frac{\Lambda }{b}+\frac{1}{2}\right)
-\psi \left( g\left( k\right) +\frac{\Lambda }{b}\right) +\psi \left(
g\left( k\right) \right) -\psi \left( g\left( k\right) +\frac{1}{2}\right) %
\right] \right.  \notag \\
&&+\left( 2g\left( k\right) -1\right) \left[ \left( \varepsilon _{b}-g\left(
k\right) -\frac{\Lambda }{b}\right) \left( \psi ^{\prime }\left( g\left(
k\right) +\frac{\Lambda }{b}+\frac{1}{2}\right) -\psi ^{\prime }\left(
g\left( k\right) +\frac{\Lambda }{b}\right) \right) \right.  \notag \\
&&\left. +\left. \left( \varepsilon _{b}-g\left( k\right) \right) \left(
\psi ^{\prime }\left( g\left( k\right) \right) -\psi ^{\prime }\left(
g\left( k\right) +\frac{1}{2}\right) \right) \right] -\frac{\Lambda b\left(
\varepsilon _{b}-1/2\right) }{\left( \Lambda +\left( g\left( k\right)
-1/2\right) b\right) ^{2}}\right \} \text{.}  \label{4-2}
\end{eqnarray}%
Combining Eq.(\ref{4-2}) and Eq.(\ref{4-1}) the intersection point $T^{\ast
}\left( H\right) $ is obtained. Detailed comparison with data follows.

\begin{table*}[tph]
\caption{Fitting parameters for $YBa_{2}Cu_{3}O_{7}$, $LaFeAsO_{0.9}F_{0.1-%
\protect \delta }$.}
\begin{center}
\renewcommand \arraystretch{1.5}
\begin{tabularx}{\textwidth}{XXXXXXXXX}
\hline
\hline
Material & $T_{c}$($K$) & $d^{\prime}(\mathring{A})$ & $H_{c2}$($T$) & $\gamma $ & $\Lambda $ & $Gi$ & $\omega$ & $k$  \\
\hline
$YBCuO$ & $87.5$ & $11.68$ & $200$ & $7.5$ & $0.30$ & $0.0011$ & $0.022$ & $1.4$ \\
$LaFeAsO$ & $20$ & $8.717$ & $50$ & $7.64$ & $0.30$ & $0.0015$ & $0.066$ & $0.7$ \\
\hline
\hline
\end{tabularx}
\end{center}
\end{table*}
\bigskip

\textit{YBCO. }Magneto - resistivity data of $YBa_{2}Cu_{3}O_{7}$ ($%
T_{c}=87.5K$) single crystal of Ref. \onlinecite{Palstra1989} is used to
analyse the intersection points of $K$. The superconducting fluctuations
component of the magneto - conductivity is obtained by subtracting the
normal part, $\sigma _{s}=1/\rho -1/\rho _{n}$. The normal state resistivity
$\rho _{n}$ in the low temperature region is given by the linear
extrapolation of the resistivity curve from the high temperature region. The
experimental data $\sigma _{s}H/H_{c2}$ as a function of temperature for
various magnetic fields of $YBa_{2}Cu_{3}O_{7}$ ($T_{c}=87.5K$) single
crystal\cite{Palstra1989} is shown in Fig.1. The experimental curves
intersect roughly at a point between $5T$ to $10T$, and $T^{\ast }=85.8K$ is
just below the critical temperature $T_{c}$. We use Eq.(\ref{epsilon}) and
Eq.(\ref{con}) to fit the data and the fitting parameters are listed in
Table I (the interlayer distance $d^{\prime }=11.68\mathring{A}$ is taken
from Ref. \onlinecite{Poole2007}). The solid curves are plotted using the
fitting parameters. The curves fit very well in the high temperature region
above $T^{\ast }=85.8K$, but the fitting in the low temperature region is
not good. The reason can be attributed to pinning, whose influence on the
conductivity becomes significant in the low temperature region.

The energy gap $\varepsilon \left( b\right) /b$ at $T^{\ast }=85.8K$ is
shown in Fig.6. The curve varies approximately from $0.2$ to $0.3$ for $5T$
to $10T$. Therefore the LLL approximation is questionable in the
intersection point region.

\begin{figure}[tph]
\begin{center}
\includegraphics[width=8cm]{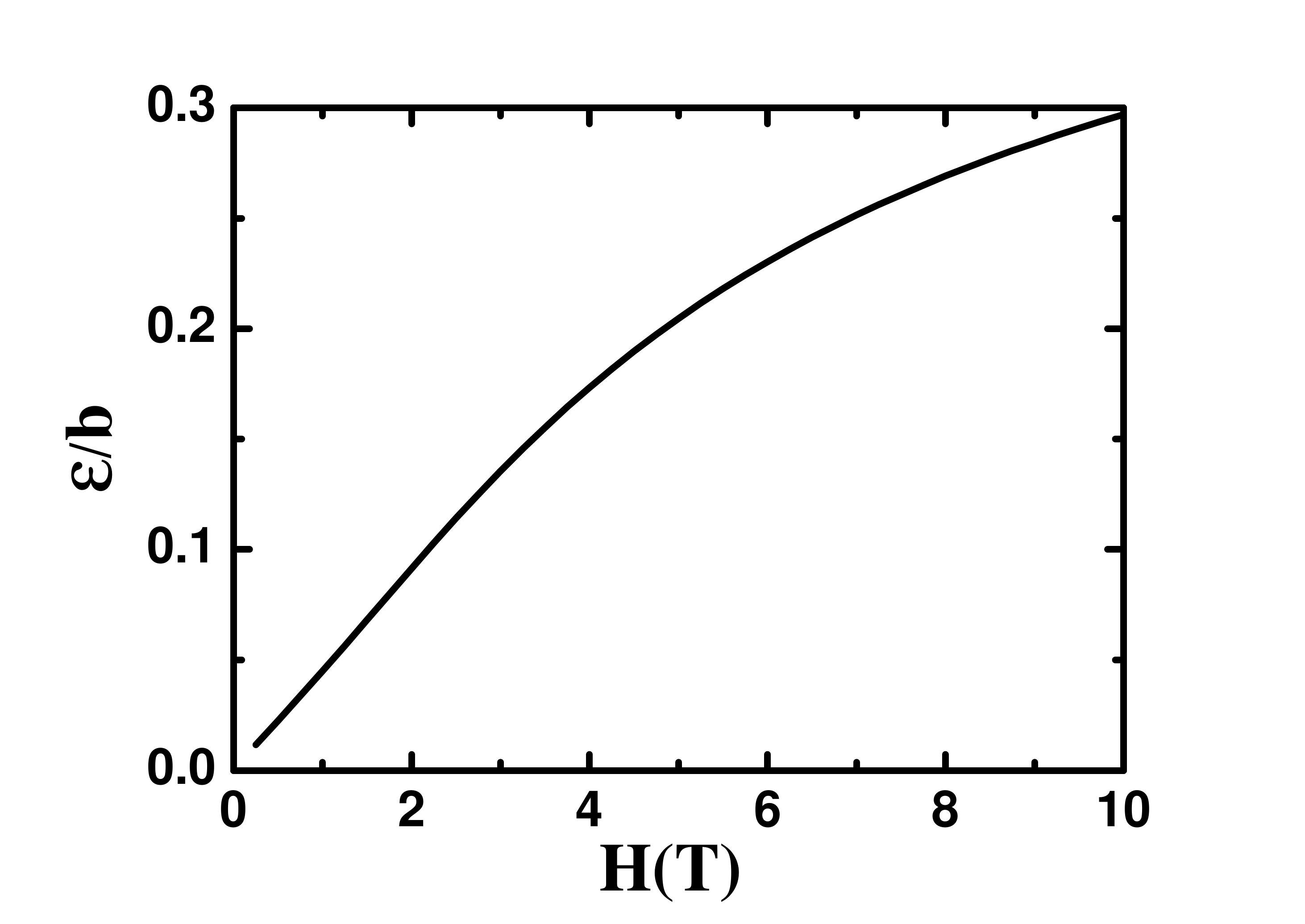}
\end{center}
\par
\vspace{-0.5cm}
\caption{ The $\protect \varepsilon \left( b\right) /b$\ as function of
magnetic field with fixed $d^{\prime }$=11.68\AA \ and $T^{\ast }=85.8$K of $%
YBCO$ crystal.}
\end{figure}
Fig.7 shows the coherence length dependence on the magnetic field at $%
T^{\ast }=85.8K$. $\xi _{c}\left( H\right) $ is quite near the interlayer
distance between $5T$ to $10T$ (good intersection point region). Hence the
appearance of intersection point is in the $2D-3D$ crossover region.

\begin{figure}[tph]
\begin{center}
\includegraphics[width=8cm]{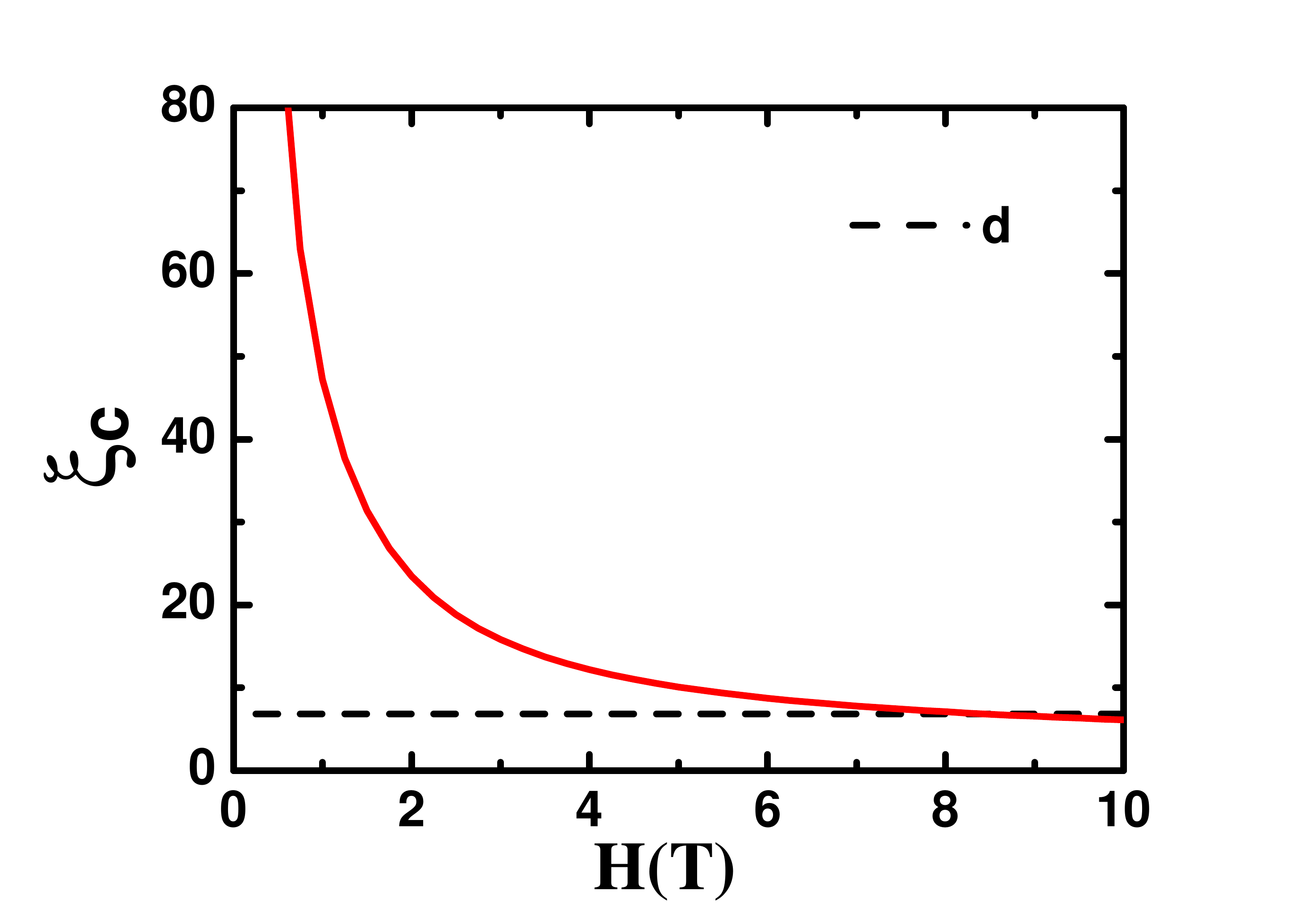}
\end{center}
\par
\vspace{-0.5cm}
\caption{$\protect \xi _{c}\left( H\right) $ as the function of magnetic
field with fixed $d^{\prime }$=11.68\AA \ and $T^{\ast }=85.8$K of $YBCO$
crystal. }
\end{figure}

Both the intersection curves for magnetization $M$ and conductivity $K$ are
plotted in Fig.8. There are quite small differences for the two curves, and
this means that the appearance of intersection points for $M$ and $K$ is due
to the same physical mechanism.
\begin{figure}[tbp]
\begin{center}
\includegraphics[width=8cm]{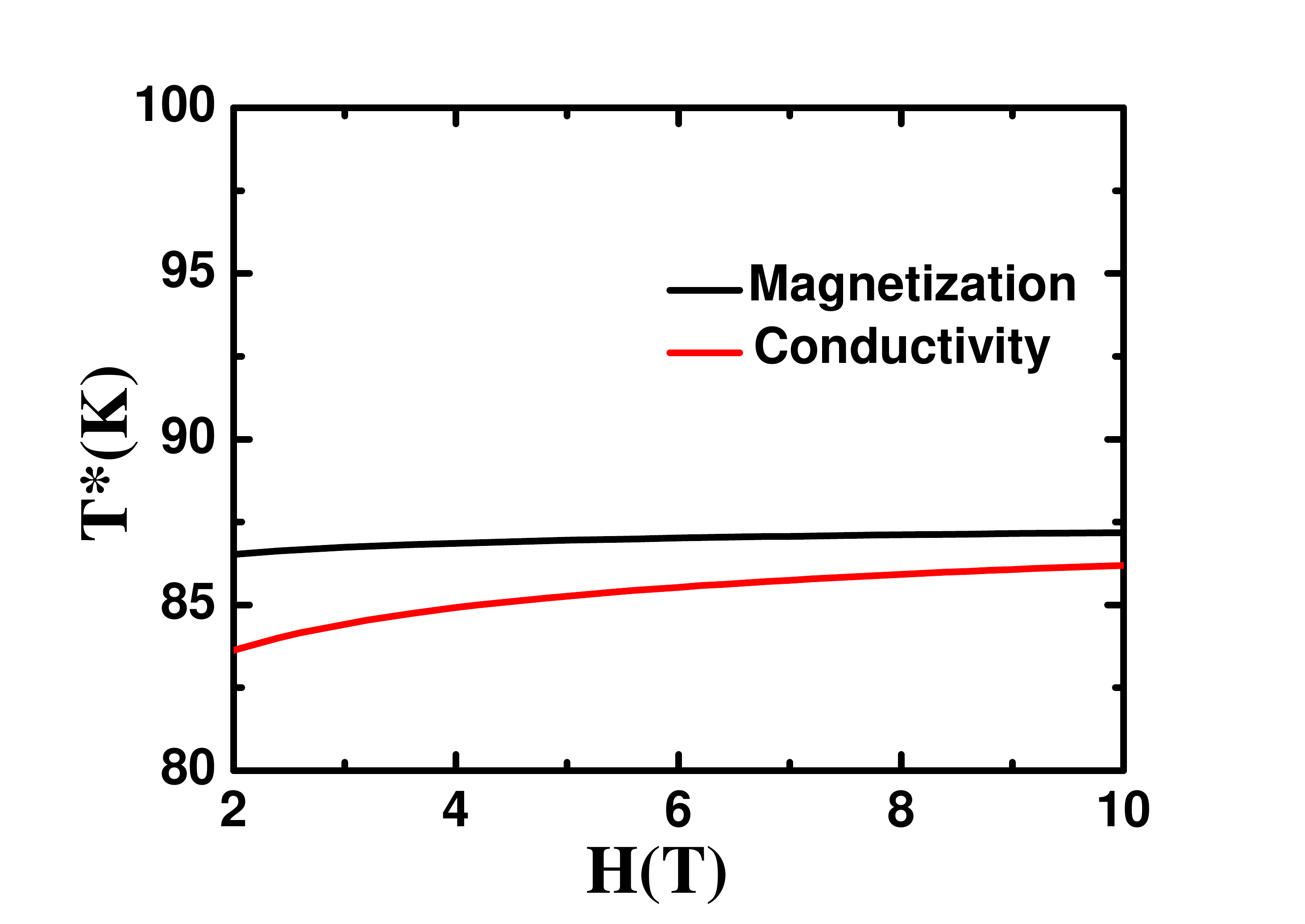}
\end{center}
\par
\vspace{-0.5cm}
\caption{The comparison of intersection point lines between magnetization
and conductivity of $YBCO$ crystal. }
\end{figure}

\textit{Pnictides. }For Fe-based superconductors, the intersection points
shall appear as they are mostly layered materials as well. The strongly
layered, $LaFeAsO_{0.9}F_{0.1-\delta }$ ($T_{c}=20K$)\cite{Chen2008} with
the anisotropy parameter $\gamma =7.64$ is considered. In Fig. 9, $\sigma
_{s}H/H_{c2}$ vs $T$ for various magnetic fields studied is shown for $%
LaFeAsO$. The fitting parameters $\gamma $, $\omega $, $H_{c2}$ had been
already established in Ref. \onlinecite{Wang2016} and are listed in Table I.

\begin{figure}[tph]
\begin{center}
\includegraphics[width=8cm]{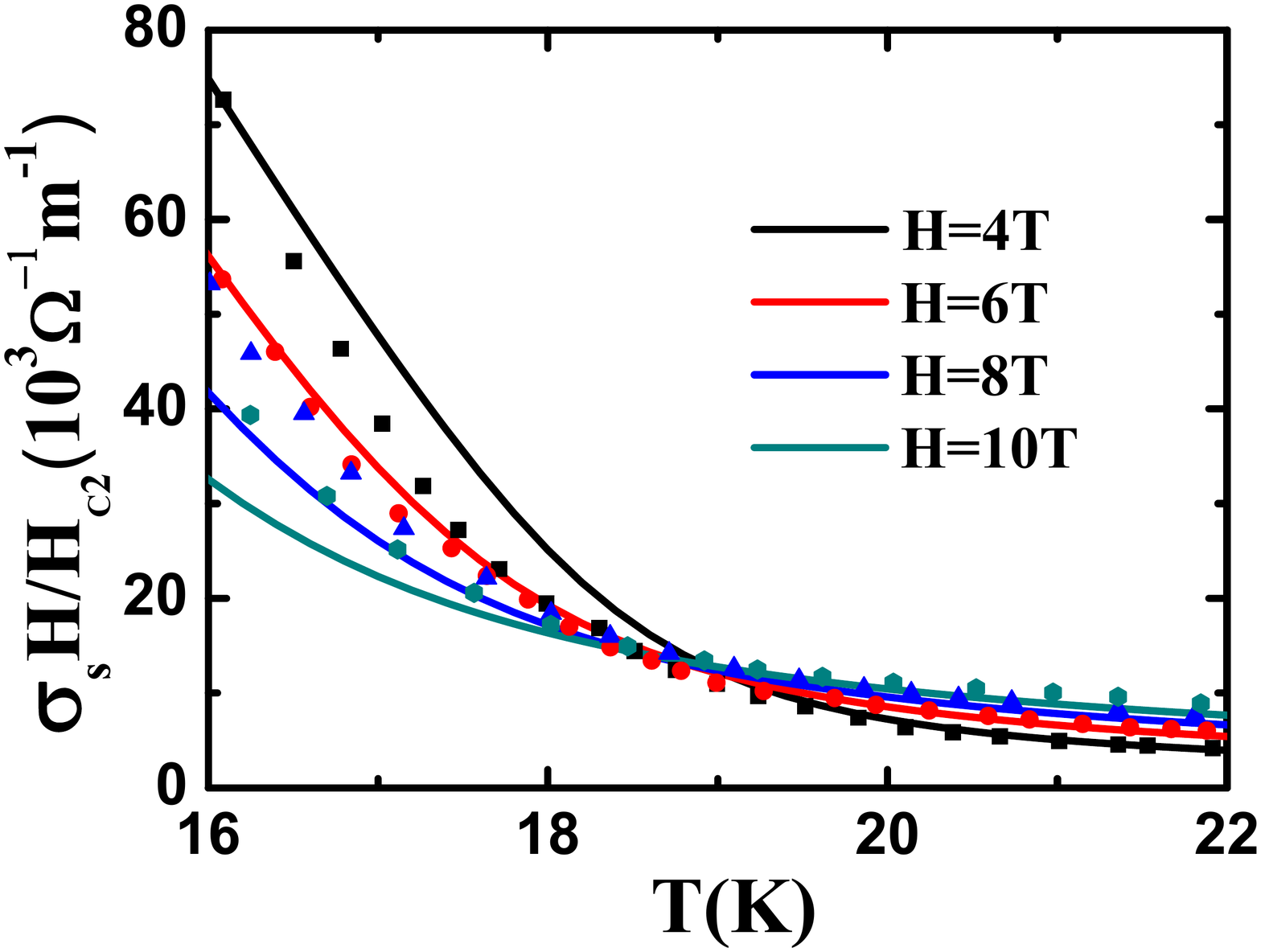}
\end{center}
\par
\vspace{-0.5cm}
\caption{Intersection points in conductivity curves of LaFeAsO$_{0.9}$F$%
_{0.1-\protect \delta }$ ($T_{c}=20$K)}
\end{figure}

\section{Summary and Conclusions}

To summarize, we noticed that in strongly fluctuating layered type II
superconductors in the vortex liquid phase,\ the product of the
superconducting part of conductivity $\sigma _{s}$ and\ of magnetic field, $%
K=\sigma _{s}H$, as a function of temperature has an ``intersection point"
similar to that noticed long ago in magnetization curves. To explain the
intersection point phenomena, the self-consistent approximation theory of
the Lawrence - Doniach model was adapted to calculate both quantities within
the same framework.

The underdoped$\ LSCO$ and optimally doped $YBCO$ high $T_{c}$ cuprates are
used as a test cases. $Fe$-based layered superconductor $%
LaFeAsO_{0.9}F_{0.1-\delta }$ \cite{Chen2008} demonstrates the existence of
the intersection point phenomenon beyond cuprates. While in the past the
``intersection points" were attributed to the 2D like superconducting
fluctuation on the basis of the lowest Landau level theory of the vortex
liquid phase\cite{Bulaevskii1992}, it was demonstrated that the LLL is not
valid near the intersection point.\ For magnetization the value of 2D-LLL
approximation is much larger than the SCFA one and these two temperature of
intersection point is not equal, so that higher Landau levels must be taken
into account.

The main observation is that the intersection point $T^{\ast }$, when it is
well defined, is located in the\ vicinity of the 2D - 3D fluctuation
crossover where the coherence length $\xi _{c}\left( T\right) $ in the
direction perpendicular to the layers is approaching the interlayer spacing.
Its location is not fixed, but changes moderately with magnetic field.

\textit{Acknowledgments.}

We are grateful to Rui Wang and Professor V. M. Krasnov for valuable
discussions. B.R. was supported by NSC of R.O.C. Grants No.
103-2112-M-009-014-MY3. The work of D.L. also is supported by National
Natural Science Foundation of China (No. 11674007). B.R. is grateful to
School of Physics of Peking University and Bar Ilan Center for
Superconductivity for hospitality.

\appendix

%%%%%%%%%%%%%%%%%%%%%%%%%%%%%%%%%%%%%%%%%%%%%%%%%%%%%%

%%%%%%%%%%%%%%%%%%%%%%%%%%%%%%%%%%%%%%%%%%%%%%%%%%%%%%

\section{2D-LLL approximation of magnetization}

Since the intersection points appear when the magnetic field is small
compared to $H_{c2}$ and temperature does not deviate too far from $T_{c}$,
let us simplify the expression, Eq.(\ref{mag}) for $b,\varepsilon \ll
\Lambda $. It takes a form%
\begin{equation}
M=-\frac{eT\gamma }{\pi hc\xi _{\Lambda }}\int_{k}\left \{ \left( g\left(
k\right) -1/2\right) \left( \psi \left( g\left( k\right) \right) -1\right)
-\ln \frac{\Gamma \left( g\left( k\right) \right) }{\sqrt{2\pi }}\right \}
\text{.}  \label{3-2}
\end{equation}%
The gap equation also simplifies:%
\begin{equation}
\varepsilon =-a_{h}-\frac{\omega t}{\pi }\left[ \ln \left( 2bd^{2}\right) +%
\frac{d}{2\pi }\int_{k}\psi \left( g\left( k\right) \right) \right] \text{.}
\label{3-3}
\end{equation}

In the LLL approximation, the inter Landau energy is much larger than the
intra Landau level excitation, that is $b\gg \varepsilon $. The
magnetization formula of Eq.(\ref{3-2}) is%
\begin{equation}
M_{LLL}=-\frac{eT}{hcd^{\prime }}\frac{b}{\sqrt{\varepsilon _{LLL}\left(
2/d^{2}+\varepsilon _{LLL}\right) }}  \label{magLLL}
\end{equation}%
with%
\begin{equation}
\varepsilon _{LLL}=-a_{h}+\frac{\omega t}{\pi }\frac{b}{\sqrt{\varepsilon
_{LLL}\left( 2/d^{2}+\varepsilon _{LLL}\right) }}
\end{equation}

By taking both the 2D and the LLL approximation, the intersection point can
be found analytically. Taking the $d\rightarrow \infty $ limit of Eq.(\ref%
{magLLL}), one obtains,
\begin{equation}
M_{LLL}^{2D}\simeq -\frac{eT}{hcd^{\prime }}\frac{b}{\varepsilon _{LLL}^{2D}}%
\text{,}  \label{3-5}
\end{equation}%
where the gap equation for $\varepsilon _{LLL}^{2D}$ within the 2D-LLL
approximation becomes just%
\begin{equation}
\varepsilon _{LLL}^{2D}\simeq -a_{h}+\frac{\omega t}{\pi }\frac{b}{%
\varepsilon _{LLL}^{2D}}\text{.}  \label{3-6}
\end{equation}%
Using these expressions, the intersection point condition, Eq.(\ref{3-1}),
allows an analytic solution:%
\begin{equation}
T_{LLL}^{2D\ast }=\frac{T_{c}}{1+4\omega /\pi }\text{.}
\end{equation}

\section{Calculation of the coherence length along the direction
perpendicular to layers.}

The interlayer coherence length $\xi _{c}$ is determined from the
exponential decrease of the order parameter correlator $\left \langle \psi
_{m}^{\ast }\psi _{n}\right \rangle $ as a function of the inter - layers
distance $\left( n-m\right) d^{\prime }$. The method correlator can be
calculated using SCFA, and we will follow Ref. \onlinecite{Jiang2014}. To
simplify the calculation, the field and length will be expressed in physical
units.

The dimensionless order parameter is $\phi =\sqrt{\beta /2\alpha T_{\Lambda }%
}\Psi ,$so that the GL Boltzmann factor in scaled units takes a form,
\begin{eqnarray}
f &=&\frac{F}{T}=\frac{1}{2\omega t}\sum_{n}\int_{\mathbf{r}}\left[ \left
\vert \mathbf{D}\phi _{n}\right \vert ^{2}+d^{-2}\left \vert \phi _{n}-\phi
_{n+1}\right \vert ^{2}\right.  \notag \\
&&\left. -\left( 1-t_{\Lambda }\right) \left \vert \phi _{n}\right \vert
^{2}+\left \vert \phi _{n}\right \vert ^{4}\right] \text{.}
\end{eqnarray}%
where the dimensionless covariant derivative $\mathbf{D}$ in the above
equation is $\left( \mathbf{\nabla +}i\left( 2e/\hbar c\right) \mathbf{A}%
\right) /\xi _{\Lambda }$. The $\phi _{n}\left( \mathbf{r}\right) $ is
expressed by Fourier transform field $\phi _{l,\mathbf{q,}k}$,%
\begin{equation}
\phi _{n}(\mathbf{r})=\frac{1}{\left( 2\pi \right) ^{3/2}}\sum_{l}\int_{%
\mathbf{q}}\int_{k}e^{indk}\varphi _{l,\mathbf{q}}(\mathbf{r})\phi _{l,%
\mathbf{q},k}\text{,}  \label{5-2}
\end{equation}%
$\varphi _{l,\mathbf{q}}(\mathbf{r})$ is the Landau's quasi - momentum wave
function\cite{Li2010}\textbf{.} The correlator is given by the statistical
average within SCFA. \
\begin{eqnarray}
\left \langle \phi _{l,\mathbf{q,}k}^{\ast }\phi _{l^{\prime },\mathbf{q}%
^{\prime }\mathbf{,}k^{\prime }}\right \rangle &=&\frac{1}{Z_{0}}\int_{\phi
}\phi _{l,\mathbf{q,}k}^{\ast }\phi _{l^{\prime },\mathbf{q}^{\prime }%
\mathbf{,}k^{\prime }}e^{-P}\text{,} \\
\text{ \  \  \ }Z_{0} &=&\int_{\phi }e^{-P}\text{,}  \notag
\end{eqnarray}%
where

\begin{equation}
P=\frac{1}{\omega td}\sum \limits_{l=0}^{\infty }\int_{\mathbf{q}}\int_{k}%
\left[ \frac{1}{d^{2}}\left( 1-\cos kd\right) +lb+\varepsilon \right] \phi
_{l,\mathbf{q},k}\phi _{l,\mathbf{q},k}^{\ast }\text{,}
\end{equation}%
leads to%
\begin{equation}
\left \langle \phi _{l,\mathbf{q,}k}^{\ast }\phi _{l^{\prime },\mathbf{q}%
^{\prime }\mathbf{,}k^{\prime }}\right \rangle =\frac{\omega td}{\left(
1-\cos kd\right) /d^{2}+lb+\varepsilon }\delta _{l,l^{\prime }}\delta \left(
\mathbf{q-q}^{\prime }\right) \delta \left( k\mathbf{-}k^{\prime }\right)
\text{.}  \label{correlatorf}
\end{equation}

The correlator between different layers is defined as
\begin{equation}
\left \langle \phi _{m}^{\ast }\left( \mathbf{r}\right) \phi _{n}\left(
\mathbf{r}\right) \right \rangle =\frac{1}{S}\int_{\mathbf{r}}\left \langle
\phi _{m}^{\ast }\left( \mathbf{r}\right) \phi _{n}\left( \mathbf{r}\right)
\right \rangle \text{,}
\end{equation}%
where $S$ is the area of the layer. Using Eq. (\ref{5-2}),
\begin{eqnarray}
&&\left \langle \phi _{m}^{\ast }\left( \mathbf{r}\right) \phi _{n}\left(
\mathbf{r}\right) \right \rangle  \notag \\
&=&\frac{1}{\left( 2\pi \right) ^{3}S}\int_{\mathbf{r}}\int_{k,k^{\prime
}}e^{-i\left( mk-nk^{\prime }\right) d}\underset{l,l^{\prime }}{\sum }\int_{%
\mathbf{q,q}^{\prime }}\mathbf{\varphi }_{l,\mathbf{q}}^{\ast }\left(
\mathbf{r}\right) \mathbf{\varphi }_{l^{\prime },\mathbf{q}^{\prime }}\left(
\mathbf{r}\right) \left \langle \phi _{l,\mathbf{q,}k}^{\ast }\phi
_{l^{\prime },\mathbf{q}^{\prime }\mathbf{,}k^{\prime }}\right \rangle
\notag \\
&=&\frac{\omega td}{\left( 2\pi \right) ^{3}S}\int_{\mathbf{r}%
}\int_{k}e^{-i\left( m-n\right) kd}\underset{l}{\sum }\int_{\mathbf{q}}%
\mathbf{\varphi }_{l,\mathbf{q}}^{\ast }\left( \mathbf{r}\right) \mathbf{%
\varphi }_{l,\mathbf{q}}\left( \mathbf{r}\right) \frac{1}{\left( 1-\cos
kd\right) /d^{2}+lb+\varepsilon }  \notag \\
&=&\frac{\omega tdb}{\left( 2\pi \right) ^{2}}\underset{l}{\sum }%
\int_{k}e^{-i\left( m-n\right) kd}\frac{1}{\left( 1-\cos kd\right)
/d^{2}+lb+\varepsilon }\text{.}
\end{eqnarray}

The result is%
\begin{equation}
\left \langle \phi _{m}^{\ast }\left( \mathbf{r}\right) \phi _{n}\left(
\mathbf{r}\right) \right \rangle =\omega t\frac{d^{2}b}{2\pi }\underset{l}{%
\sum }\frac{\left( Q_{l}-\sqrt{Q_{l}^{2}-1}\right) ^{n-m}}{\sqrt{Q_{l}^{2}-1}%
}\text{,}
\end{equation}%
in which $Q_{l}=\left( lb+\varepsilon \right) d^{2}+1$. For $n-m>>1$, the
biggest value of $Q_{l}-\sqrt{Q_{l}^{2}-1}$survives, that happens when $l=0$,%
\begin{equation}
\left \langle \phi _{m}^{\ast }\left( \mathbf{r}\right) \phi _{n}\left(
\mathbf{r}\right) \right \rangle _{n-m>>1}\simeq \omega t\frac{d^{2}b}{2\pi }%
\frac{\exp \left[ -\frac{\left( n-m\right) d}{\xi _{c}}\right] }{\sqrt{%
\left( \varepsilon d^{2}+1\right) ^{2}-1}}\text{,}
\end{equation}%
so the coherence length (in unit $\xi _{\Lambda }/\gamma $) is%
\begin{equation}
\xi _{c}=-\frac{d}{\ln \left( \varepsilon d^{2}+1-\sqrt{\left( \varepsilon
d^{2}+1\right) ^{2}-1}\right) }\text{.}
\end{equation}%
This was used in Fig. 5 and Fig. 7.

When magnetization was measured precisely enough and the normal background
carefully subtracted\cite{LuLi2010}, it was found surprisingly that, when
the magnetization as a function of temperature, $M\left( T\right) $, plotted
at different magnetic fields $H$, the curves intersect at the same
temperature $T^{\ast }$.



\begin{thebibliography}{99}
\bibitem{Zeldov1995} E. Zeldov, D. Majer, M. Konczykowski, V. B.
Geshkenbein, V. M. Vinokur, and H. Shtrikman, Nature (London) \textbf{375},
373 (1995).

\bibitem{Schilling1997} A. Schilling, R.A. Fisher, N.E. Phillips, U. Welp,
W.K. Kwok, and G.W. Crabtree, Phys. Rev. Lett. \textbf{78}, 4833 (1997).

\bibitem{Junod1999} A. Junod, A. Erb, and C. Renner, Physica C \textbf{317},
333 (1999).

\bibitem{LuLi2010} L. Li, Y. Wang, S. Komiya, S. Ono, Y. Ando, G. D. Gu, and
N. P. Ong, Phys. Rev. B \textbf{81}, 054510 (2010).

\bibitem{Rullier2011} F. Rullier-Albenque, H. Alloul, and G. Rikken, Phys.
Rev. B \textbf{84}, 014522 (2011).

\bibitem{Xu2000} Z. A. Xu, N. P. Ong, Y. Wang, T. Kakeshita and S. Uchida,
Nature \textbf{406}, 486 (2000).

\bibitem{Wang2006} Y. Wang, L. Li, and N. P. Ong, Phys. Rev. B \textbf{73},
024510 (2006).

\bibitem{Welp1991} U. Welp, S. Fleshler, W. K. Kwok, R. A. Klemm, V. M.
Vinokur, J. Downey, B. Veal, and G. W. Crabtree, Phys. Rev. Lett. \textbf{67}%
, 3180 (1991).

\bibitem{Kes1991} P.H. Kes, C. J. van der Beek, M. P. Maley, M. E. McHenry,
D. A. Huse, M. J. V. Menken, and A. A. Menovsky, Phys. Rev. Lett. \textbf{67}%
, 2383 (1991).

\bibitem{Wahl1997} A. Wahl, V. Hardy, F. Warmont, A. Maignan, M. P.
Delamare, and C. Simon, Phys. Rev. B \textbf{55}, 3929 (1997).

\bibitem{Naughton2000} M. J. Naughton, Phys. Rev. B, \textbf{61}, 1605
(2000).

\bibitem{Finnemore2002} Y. M. Huh and D. K. Finnemore, Phys. Rev. B \textbf{%
65}, 092506 (2002).

\bibitem{Thouless1975} D. J. Thouless, Phys. Rev. Lett. \textbf{34}, 946
(1975).

\bibitem{Thouless1976} G. J. Ruggeri and D. J. Thouless, J. Phys. F: Met.
Phys. \textbf{6}, 2063 (1976).

\bibitem{Tesanovic1994} Z. Te\v{s}anovi\'{c} and A. V. Andreev, Phys. Rev. B
\textbf{49}, 4064 (1994).

\bibitem{Bulaevskii1992} Z. Te\v{s}anovi\'{c}, L. Xing, L. Bulaevskii, Q.
Li, and M. Suenaga, Phys. Rev. Lett. \textbf{69}, 3563 (1992).

\bibitem{Rosenstein2005} FarehPei-Jen Lin and B. Rosenstein, Phys. Rev. B
\textbf{71}, 172504 (2005).

\bibitem{Jiang2014} X. Jiang, D. Li and B. Rosenstein, Phys. Rev. B \textbf{%
89}, 064507 (2014).

\bibitem{Dorsey1991} S. Ullah and A. T. Dorsey, Phys. Rev. Lett. \textbf{65}%
, 2066 (1990); Phys. Rev. B \textbf{44}, 262 (1991). \

\bibitem{Lan1993} M. D. Lan, J. Z. Liu, Y. X. Jia, Lu Zhang, Y. Nagata, P.
Klavins, and R. N. Shelton, Phys. Rev. B \textbf{47}, 457 (1993).

\bibitem{Kim1992} D. H. Kim, K. E. Gray, and M. D. Trochet, Phys. Rev. B
\textbf{45}, 10801 (1992).

\bibitem{Pallecchi2009} I. Pallecchi, C. Fanciulli, M. Tropeano, A.
Palenzona, M. Ferretti, A. Malagoli, A. Martinelli, I. Sheikin, M. Putti,
and C. Ferdeghini, Phys. Rev. B \textbf{79}, 104515 (2009)

\bibitem{Liu2010} S. L. Liu, W. Haiyun, and B. Gang, Phys. Lett. A \textbf{%
374}, 3529 (2010).

\bibitem{Palstra1989} T. T. M. Palstra, B. Batlogg, R. B. van Dover, L. F.
Schneemeyer, and J. V. Waszczak, Appl. Phys. Lett. \textbf{54}, 763 (1989).

\bibitem{Chen2008} G.F. Chen, Z. Li, G. Li, J. Zhou, D. Wu, J. Dong, W.Z.
Hu, P. Zheng,Z.J. Chen, H.Q. Yuan, J. Singleton, J.L. Luo, and N.L. Wang,
Phys. Rev. Lett. \textbf{101}, 057007 (2008).

\bibitem{Hohenberg1977} P. C. Hohenberg and B. I. Halperin, Rev. Mod. Phys.
\textbf{49}, 435 (1977).

\bibitem{Bui2010} B. D. Tinh, D. Li, and B. Rosenstein, Phys. Rev. B \textbf{%
81}, 224521 (2010).

\bibitem{Wang2016} R. Wang and D. Li, Chin. Phys. B \textbf{25}, 097401
(2016).

\bibitem{Salem09} S. Salem-Sugui, Jr., J. Mosqueira, and A. D. Alvarenga,
Phys. Rev. B \textbf{80}, 094520 (2009); S. Salem-Sugui Jr., A. D.
Alvarenga, J. Mosqueira, J. D. Dancausa, C. Salazar Mejia, E. Sinnecker1, H.
Luo and H. Wen, Supercond. Sci. Technol. \textbf{25}. 105004 (2012); J.
Mosqueira, L. Cabo, and F. Vidal, Phys. Rev. B \textbf{76}, 064521 (2007).

\bibitem{Rosenstein2001} B. Rosenstein, B. Ya. Shapiro, R. Prozorov, A.
Shaulov, and Y. Yeshurun, Phys. Rev. B \textbf{63}, 134501 (2001).

\bibitem{Poole2007} C. P. Poole Jr., H. A. Farach, R. J. Creswick, and R.
Prozorov, {\itshape Superconductivity} (Academic Press, Amsterdam, 2007).

\bibitem{Li2010} B. Rosenstein and D. Li, Rev. Mod. Phys. \textbf{82}, 109
(2010).\newline
\end{thebibliography}
\end{document}